\documentclass[runningheads]{llncs}

\usepackage[dvipsnames]{xcolor}
\usepackage{graphicx}
\usepackage{acro}
\usepackage{amsmath}
\usepackage{amssymb}
\usepackage{graphicx}
\usepackage{epsfig}
\usepackage{graphicx}
\usepackage{amsmath}
\usepackage{amssymb}
\usepackage{bm}
\usepackage{makecell}
\usepackage{mathtools}
\usepackage{pifont}
\usepackage{framed,multirow}
\usepackage{threeparttable}

\usepackage{xcolor}
\definecolor{citecolor}{HTML}{0071BC}
\definecolor{linkcolor}{HTML}{ED1C24}
\usepackage[colorlinks,
            anchorcolor=red,
            citecolor=citecolor, 
            linkcolor=linkcolor,
            ]{hyperref}

\makeatletter

\newcommand\footnoteref[1]{\protected@xdef\@thefnmark{\ref{#1}}\@footnotemark}
\makeatother

\newcolumntype{P}[1]{>{\centering\arraybackslash}p{#1}}
\newlength\savewidth\newcommand\shline{\noalign{\global\savewidth\arrayrulewidth
  \global\arrayrulewidth 0.8pt}\hline\noalign{\global\arrayrulewidth\savewidth}}

\newcommand{\etal}{\mbox{et al.}}
\newcommand{\ie}{\mbox{i.e.,\ }}

\def\arrvline{\hfil\kern\arraycolsep\vline\kern-\arraycolsep\hfilneg}

\begin{document}
%
\title{Early Detection and Localization of Pancreatic Cancer by Label-Free Tumor Synthesis}
\titlerunning{Label-Free Pancreatic Tumor Synthesis}

\author{Bowen Li\inst{1} \and
Yu-Cheng Chou\inst{1} \and
Shuwen Sun\inst{2} \and
Hualin Qiao\inst{3} \\
Alan Yuille\inst{1} \and
Zongwei Zhou\inst{1,}\thanks{Corresponding author: Zongwei Zhou (\href{mailto:zzhou82@jh.edu}{zzhou82@jh.edu})}
}
\authorrunning{B. Li et al.}
%
\institute{Johns Hopkins University \and
The First Affiliated Hospital of Nanjing Medical University \and
Rutgers University \\
{\scriptsize Code:~\href{https://github.com/MrGiovanni/SyntheticTumors}{https://github.com/MrGiovanni/SyntheticTumors}}
}

\maketitle              
\begin{abstract}

Early detection and localization of pancreatic cancer can increase the 5-year survival rate for patients from 8.5\% to 20\%. Artificial intelligence (AI) can potentially assist radiologists in detecting pancreatic tumors at an early stage. Training AI models require a vast number of annotated examples, but the availability of CT scans obtaining early-stage tumors is constrained. This is because early-stage tumors may not cause any symptoms, which can delay detection, and the tumors are relatively small and may be almost invisible to human eyes on CT scans. To address this issue, we develop a tumor synthesis method that can synthesize enormous examples of small pancreatic tumors in the healthy pancreas without the need for manual annotation. Our experiments demonstrate that the overall detection rate of pancreatic tumors, measured by Sensitivity and Specificity, achieved by AI trained on synthetic tumors is comparable to that of real tumors. More importantly, our method shows a much higher detection rate for small tumors. We further investigate the per-voxel segmentation performance of pancreatic tumors if AI is trained on a combination of CT scans with synthetic tumors and CT scans with annotated large tumors at an advanced stage. Finally, we show that synthetic tumors improve AI generalizability in tumor detection and localization when processing CT scans from different hospitals. Overall, our proposed tumor synthesis method has immense potential to improve the early detection of pancreatic cancer, leading to better patient outcomes.

\end{abstract}
\section{Introduction}
\label{sec:introduction}

\begin{figure*}
  \centering
    \includegraphics[width=\linewidth]{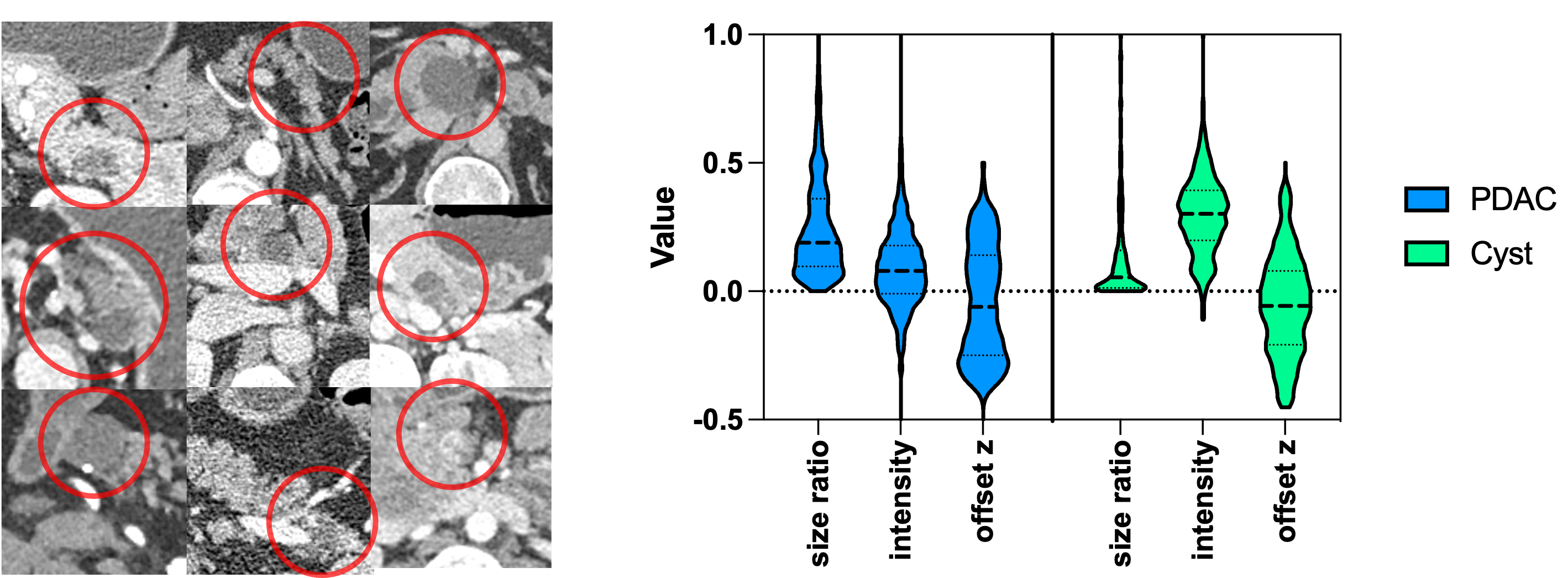}
    \caption{Visualizations of synthetic and real pancreatic tumors and real tumor statistics. 
    Left panel: \textit{Could you tell which tumors are real?} 
    Right panel: Selected normalized statistics of PDAC and Cyst, including size ratio (tumor/pancreas), normalized intensity residual (pancreas-tumor), and offset z of centers of the tumor and pancreas.}
    \label{fig:visualization-statistics}
\end{figure*}

Early detection and precise localization of cancerous tumors are crucial for disease diagnosis, prognosis, radiotherapy planning, and surgical intervention. Pancreatic cancer, one of the deadliest forms of cancer, can be detected by visual inspection of computed tomography (CT) scans, but early signs can be subtle and easily overlooked by radiologists unless the cancer is already suspected~\cite{uhm2021deep,chu2021pancreatic}. However, there are no grounds for suspicion for the vast majority of the 40 million abdominal CT scans taken annually in the United States; hence, pancreatic cancer may not be detected promptly. Artificial intelligence (AI) has the potential to assist radiologists in the early detection of cancer~\cite{xia2022felix,liu2023clip,zhang2023continual}. Although AI technology is constantly innovating, AI is data-hungry and requires large-scale, high-quality annotated datasets. Annotating early-stage pancreatic tumors is costly, time-consuming, and requires deep medical expertise, radiology reports, and biopsy results for precise annotation~\cite{zhou2021towards,zhou2022interpreting}.

To address this challenge, we synthesize PDACs (pancreatic ductal adenocarcinoma) and Cysts in the pancreas, including small tumors that are critical for early detection and diagnosis, to train AI to detect and localize tumors. Synthetic tumors refer to artificially generated tumors that are added to CT scans using computer algorithms~\cite{yao2021label,hu2022synthetic,lyu2022pseudo,wei2022pancreatic,wang2022anomaly,zhang2023self}. Synthetic tumors can be used to generate large datasets of paired CT scans and tumor masks without the need for human annotation. The synthetic tumors are generated using a model-based approach (detailed in \S\ref{sec:method}) that simulates the appearance of real tumors. The resulting dataset of synthetic pancreatic tumors can be used to train AI models for the early detection of real pancreatic tumors. Three datasets are used for evaluation, \ie \texttt{MSD}-Pancreas~\cite{antonelli2021medical}, \texttt{PCL}~\cite{abel2021automated}, and \texttt{JHH}~\cite{xia2022felix} (private data). The results show that AI trained on synthetic tumors is comparable in the overall detection rate of pancreatic tumors to those trained on real tumors. Importantly, the use of synthetic tumors led to a significantly higher detection rate for small tumors, which are particularly challenging to detect using CT scans of real tumors (due to the lack of sufficient examples)~\cite{kenner2021artificial}. Moreover, in conjunction with enormous synthetic tumors, training AI using CT scans with annotated tumors at an advanced stage (\ie large tumors) leads to more accurate segmentation of pancreatic tumors than AI trained solely on real tumors.
Finally, synthetic tumors can help overcome the problem of dataset bias, improving AI generalizability to CT scans collected from a variety of hospitals.
We summarize the contributions of this paper as follows:

\begin{enumerate}
    \item We develop an effective data synthesis method for pancreatic tumors, overcoming several technical challenges such as tumor location, intensity distribution, and texture characteristics (\figureautorefname~\ref{fig:visualization-statistics}).
    
    \item We quantify the fidelity of synthetic tumors by conducting reader studies on two radiologists and training a binary real/fake classifier. In differentiating between real and synthetic tumors, radiologists and the binary classifier yield an accuracy of 70.0\% and 62.1\%, respectively.
   (\figureautorefname~\ref{fig:roc_human_study}).

    \item AI trained on synthetic tumors achieves comparable detection performance to AI trained on real tumors. With fewer than one false positive per subject, Cyst detection could achieve an average Sensitivity of 95.8\%, and PDAC detection could achieve an average Sensitivity of 95.6\% (\tableautorefname~\ref{tab1e-FROC}).

    \item We show the ability of early tumor detection for our synthetic data strategy on our \texttt{JHH} dataset. For early-stage pancreatic tumors (radius $<$ 20mm), the model trained with synthetic data performed slightly better than the model trained with real data (\figureautorefname~\ref{fig:detection}).
    
    \item We demonstrate the improved performance using both real and synthetic tumors for training. AI trained on a small set of real tumors (around 100 cases) and enormous synthetic tumors performs significantly better than that trained solely on real tumors, highlighting the potential of synthetic tumors as complementary data augmentation.
    (\tableautorefname~\ref{tab1e-hybrid})
 
    \item  We demonstrate the effectiveness of synthetic tumors for domain adaptation, wherein the AI model was trained on public datasets and can be generalized to our in-house \texttt{JHH} dataset without additional annotation (\tableautorefname~\ref{tab1e-ablation}).

\end{enumerate}

\noindent\textbf{Related work.} By synthesizing images of diseased regions, researchers can generate large datasets that can be used to train AI algorithms for disease detection and segmentation in many organs, including colon~\cite{shin2018abnormal}, lung~\cite{yao2021label,lyu2022pseudo}, brain~\cite{wyatt2022anoddpm}, and others~\cite{wang2022anomaly,horvath2022metgan,zhang2023self}. However, the performance of AI trained on these synthetic data was significantly worse than that trained on real data. Most recently, Hu~\etal~\cite{hu2022synthetic,hu2023label} devised a holistic tumor synthesis method that suggests AI trained on synthetic data can perform similarly to real data. But this conclusion has not been generalized to more sophisticated organs other than the liver. This paper focuses on synthesizing two subtypes of tumors in the pancreas, which are much harder to synthesize due to several factors. 

\begin{itemize}
    \item The pancreas is a relatively small organ surrounded by other structures, such as the liver, stomach, and intestines, making it difficult to visualize and isolate tumors in CT scans.

    \item Pancreatic tumors have complex and heterogeneous morphology, making it difficult to capture the full range of tumor features using synthetic data.

    \item Early-stage tumors are rare, small, and difficult to detect in CT scans, making it challenging to collect a sufficiently large number of examples. 
\end{itemize}

\begin{figure}[t]
\centerline{\includegraphics[width=1.0\columnwidth]{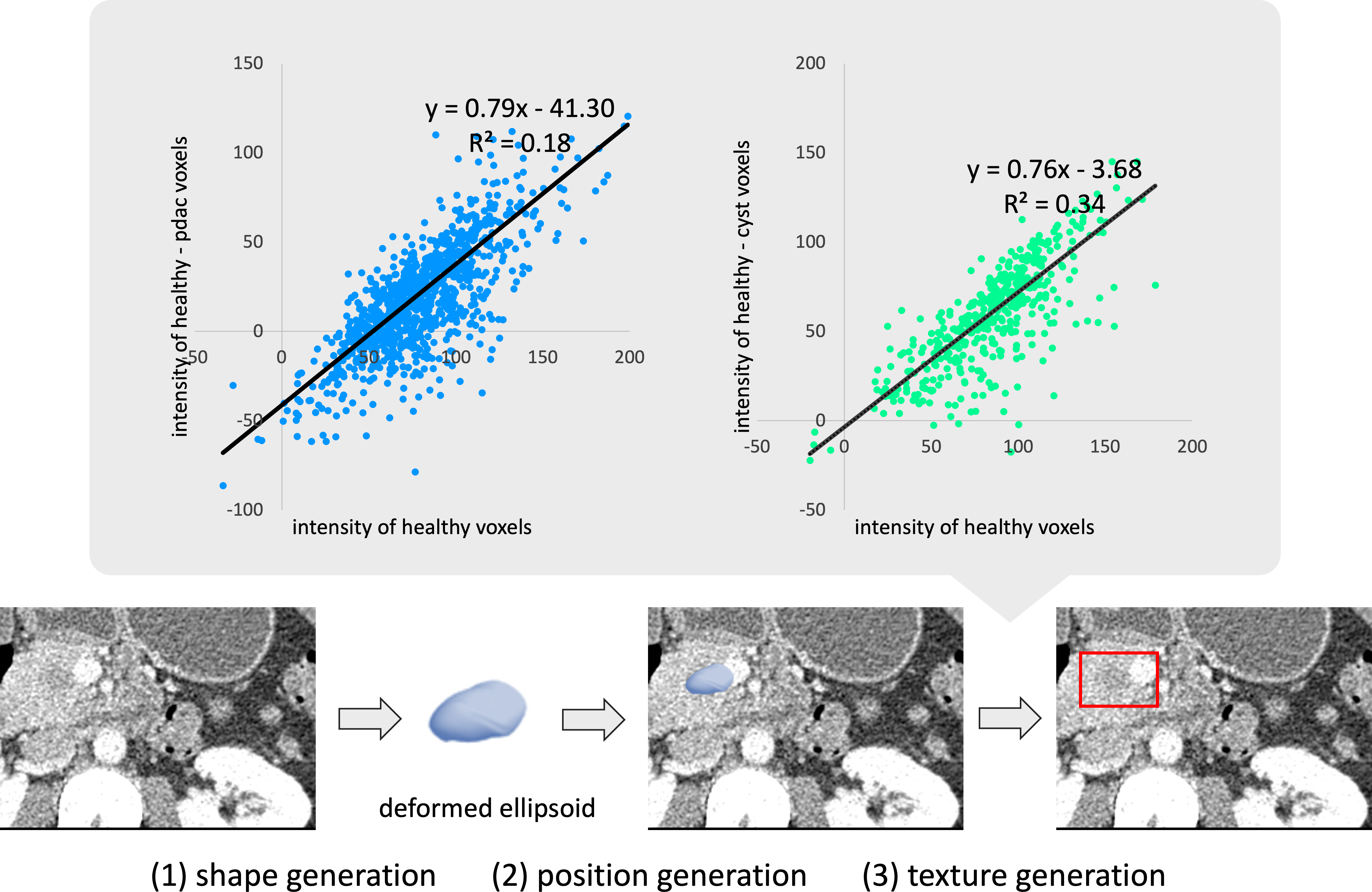}}
    \caption{
    The tumor synthesis strategy comprises three steps, all of which are based on the statistics depicted in  \figureautorefname~\ref{fig:visualization-statistics}. In addition, the texture generation step takes into account the intrinsic correlation between healthy and cancerous voxel intensities.
    }
\label{fig:tumor-generation}
\end{figure}

\section{Method}
\label{sec:method}

\figureautorefname~\ref{fig:tumor-generation} depicts our proposed tumor\footnote{In the method section, we use the term \textit{tumors} to refer to both PDACs and Cysts, as they are synthesized using the same tumor generation steps.} synthesis strategy, which contains shape, position, and texture generation. All three steps are built based on the statistics of pancreatic tumors in the \texttt{JHH} dataset~\cite{xia2022felix} as shown in \figureautorefname~\ref{fig:visualization-statistics} and \figureautorefname~\ref{fig:tumor-generation}.

\smallskip\noindent\textbf{Shape generation.}
During the statistical analysis phase, we calculated the bounding boxes of the pancreatic tumors and examined their shapes. Our analysis revealed that ellipsoids provided the best fit for pancreatic tumors, although some modifications were necessary to improve accuracy. To obtain the dimensions of the ellipsoid, we sampled data from the distribution of tumor sizes, as depicted in the right panel of \figureautorefname~\ref{fig:visualization-statistics} (size ratio). The tumor size distribution followed a positively skewed normal distribution. To incorporate more small tumors into our model, we manually stratified the sizes and introduced additional smaller tumors. Next, we applied an elastic deformation to the target ellipsoid to achieve a more natural shape.

\smallskip\noindent\textbf{Position generation.}
Once the tumor shape is defined, its position needs to be determined. As shown in the right panel of \figureautorefname~\ref{fig:visualization-statistics} (offset z), pancreatic tumors may manifest anywhere within the pancreas, from the head to the tail. Therefore, it is sensible to randomly select a point within the pancreas as the center of the synthetic tumor. Such a positioning strategy, however, could result in a synthetic tumor where most of its content is outside the pancreas. While such a scenario is plausible in the human body, managing these extreme cases during the texture generation phase is imperative.

\smallskip\noindent\textbf{Texture generation.}
For each generated tumor $T$ with a fixed position $x,y,z$, a small neighborhood of pancreas $N(T_{x,y,z}, r)$ around the tumor was sampled to calculate the median intensity difference of the tumor $\Delta I$, where $r$ is the sampling radius:
$$
\Delta I = \alpha \cdot(N(T_{x,y,z},r) \times P)+\beta+\epsilon
$$
$\alpha, \beta$ are linear regression coefficients of intensity (see \figureautorefname~\ref{fig:tumor-generation}), and $\epsilon$ is a small random term to add more randomness to intensity. The texture of the tumor follows a normal distribution. Finally, a Gaussian blur mask is applied to the generated texture to smooth the edge of the tumor. For tumors with extreme positions, only pancreas intensity (instead of all surrounding organs) should be calculated for $\Delta I$.

\section{Experiment \& Result}
\label{sec:experiment_result}


\smallskip\noindent\textbf{Implementation and datasets.}
We tested our algorithm on three datasets. Our in-house dataset \texttt{JHH} is a contrast-enhanced pancreatic CT dataset. For training, \texttt{JHH} has 662 healthy cases, 443 cases with Cyst, and 1,015 cases with PDAC. For testing purposes, there are 323 cases which is a hybrid of healthy, Cyst, and PDAC cases. Two public datasets were included during the study: \texttt{MSD} and \texttt{PCL}. Because the testing labels were not available to us, we only used images from the training set. We randomly split \texttt{MSD} to 197 in the training set and 84 in the test set. We randomly split \texttt{PCL} to 152 in the training set and 52 in the test set. All cases in the \texttt{MSD} dataset are diseased, but whether the patients had Cysts or PDACs was unidentified. All cases in \texttt{PCL} were verified to contain Cysts.
To calculate the statistics of pancreatic tumors, 100 cases with PDAC and 100 cases with Cyst were randomly sampled from \texttt{JHH}. 
For the Turing Visual Test, we randomly selected 50 cases with Cyst and 50 healthy cases with synthetic Cyst from \texttt{JHH}. For each case, one slice with the tumor was extracted, together with the tumor masks. 
During the training stage, synthetic tumors were generated on the fly. During each epoch, each healthy image yields two images with synthetic tumors. The tumor's size and position were randomly sampled on the tumor statistics, and the median intensity of the tumor was calculated with the intensity regression.

\begin{figure}[t]
\centerline{\includegraphics[width=1.0\columnwidth]{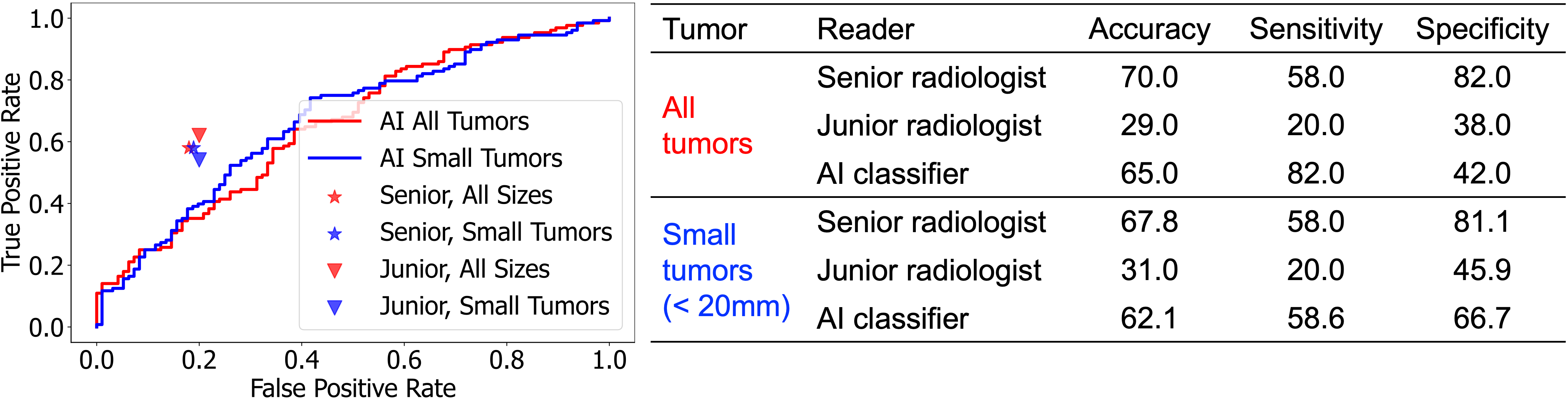}}
    \caption{
    The ROC curves were plotted to evaluate the performance of an AI binary classifier and two radiologists (a junior with six months of experience and a senior with five years of experience). To conduct the study, we generated 50 synthetic tumors and mixed them with 50 real tumors. The resulting set of scans was presented to the radiologists for testing without disclosing the ratio of real to synthetic tumors in the dataset. The synthetic tumors generated by our method are realistic and could confound human readers. Some examples of the Visual Turing Test are in \figureautorefname~\ref{fig:visualization-statistics}. 
    }
\label{fig:roc_human_study}
\end{figure}

\begin{table*}[t]
\caption{
AI trained by synthetic tumors achieves an average Sensitivity improvement from 81.3\% to 97.8\% at 0.7, 0.8, 0.9, and 1.0 false positives per subject compared with AI trained by real tumors; however, synthetic tumors are outperformed by real tumors for segmenting pancreatic cysts. In general, label-free synthetic tumors enable AI to segment pancreatic tumors as effectively as label-intensive real tumors.
}
\centering
\scriptsize
\begin{tabular}{p{0.1\linewidth}P{0.2\linewidth}|P{0.1\linewidth}P{0.1\linewidth}P{0.1\linewidth}P{0.1\linewidth}P{0.15\linewidth}}
    \hline
    tumor & training data &$0.7$&$0.8$&$0.9$&$1.0$ & average \\
    \shline
    \multirow{2}{*}{PDAC} & synt &$96.8$&$97.4$&$98.5$&$98.8$&$97.8$\\
     & real &$70.2$&$79.5$&$83.9$&$91.4$&$81.3$\\
    \multirow{2}{*}{Cyst} & synt &$98.4$&$98.7$&$97.7$&$91.4$&$96.6$\\
     & real &$97.5$&$98.6$&$99.1$&$99.8$&$98.7$\\
    \hline
\end{tabular}
\label{tab1e-FROC}
\end{table*}

\begin{table*}[t]
\caption{
AI trained by a combination of real and synthetic tumors performs better than that trained solely on real tumors. This has been validated on two publicly available datasets, suggesting that synthetic tumors have the potential to serve as a promising data augmentation strategy as complementary to real tumors. Therefore, numerous public CT scans with a healthy pancreas---relatively easier to collect with the help of radiology reports---can also be augmented and used for training AI models for tumor segmentation.
}
\centering
\scriptsize
\begin{tabular}{p{0.18\linewidth}p{0.15\linewidth}p{0.15\linewidth}|P{0.1\linewidth}P{0.35\linewidth}}
\hline
tumor & training set & test set &  DSC & Sensitivity @0.05FP/subject \\
\shline
real & \texttt{PCL} & \texttt{PCL} &$50.9$ & $98.8$\\
real + synt & \texttt{PCL} & \texttt{PCL} &$51.8$ & $100.0$\\
\hline
real & \texttt{MSD} & \texttt{MSD}
    &$41.9$ & $97.0$\\

real + synt & \texttt{MSD} & \texttt{MSD}
    &$43.2$ & $97.0$\\
\hline
\end{tabular}
\label{tab1e-hybrid}
\end{table*}

\smallskip\noindent\textbf{Results of Visual Turing Test.} 
First, we need to understand whether the synthetic tumors generated look realistic or not.  The capability of synthesizing realistic tumors is important because it enables us to generate a lot of training data for AI algorithms, particularly to get examples of small tumors that may appear only occasionally in the annotated datasets. To assess tumor synthesis quality visually, we recruited two radiologists to distinguish ``fake'' tumors generated by our method from real tumors. Each radiologist was provided with 100 images, and half of them were synthesized.
As \figureautorefname~\ref{fig:roc_human_study} shows, for tumors of all sizes, the senior radiologist could correctly classify 70\% of the images. However, the junior radiologist could only correctly identify 29\% of them. The results suggest that about 60\% of synthetic data could even fool a clinician with domain expertise. We further stratified the results by the radius of the tumor (the threshold was 20mm as recommended by our radiologists), and the classification accuracy of the senior radiologist dropped, which coincides with our assumption: small tumors are harder to observe and are more difficult to diagnose. 

\begin{figure*}[t]
  \centering
    \includegraphics[width=\linewidth]{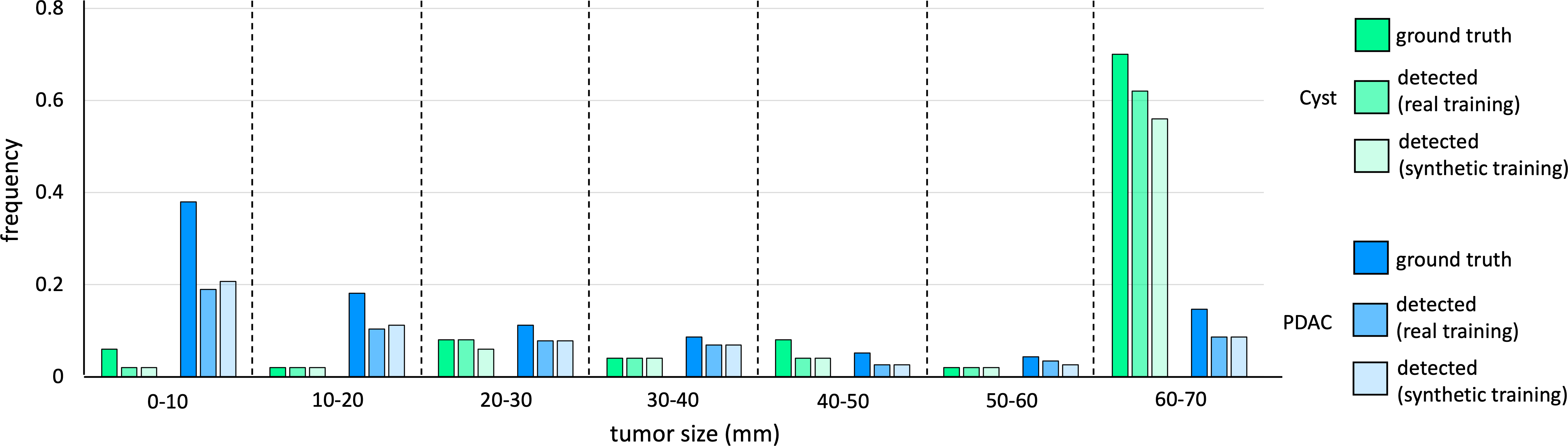}
  \caption{AI trained on synthetic tumors smaller than 20mm outperforms that trained on real tumors due to difficulty in collecting and annotating CT scans with small tumors. The lack of training examples limits AI performance in small tumor detection. In contrast, synthetic tumors provide diverse and ample examples of small tumors, therefore improving AI performance.}
  \label{fig:detection}
\end{figure*}

\begin{table*}[t]
\caption{
Synthetic tumors enable more robust domain adaptation because they are generated to CT scans collected from novel domains (hospitals). We evaluate this by using our large-scale in-house dataset, namely \texttt{JHH}, where PDAC and Cyst are annotated. The AI model is trained on  CT scans from public datasets with real tumors and CT scans from the in-house dataset with a healthy pancreas. The Sensitivity (Sen) is computed at 0.7 false positives per subject.
}
\centering
\scriptsize
\begin{tabular}{p{0.15\linewidth}p{0.15\linewidth}p{0.15\linewidth}|P{0.1\linewidth}P{0.14\linewidth}|P{0.1\linewidth}P{0.14\linewidth}}
\hline
 & & &\multicolumn{2}{c|}{PDAC} &\multicolumn{2}{c}{Cyst}\\
tumor & training set & test set &  DSC & Sensitivity & DSC & Sensitivity \\
\shline
real & \texttt{PCL} & \texttt{JHH}
    &$25.6$ & $76.8$
    &$40.2$ & $95.7$\\
real + synt & \texttt{PCL} & \texttt{JHH}
    &$30.2$ & $74.0$
    &$45.2$ & $94.3$\\
\hline
real & \texttt{MSD} & \texttt{JHH}
    &$51.2$ & $79.7$
    &$54.1$ & $96.5$\\
real + synt & \texttt{MSD} & \texttt{JHH}
    &$52.2$ & $87.6$
    &$57.2$ & $97.9$\\
\hline
\end{tabular}
\label{tab1e-ablation}
\end{table*}

\smallskip\noindent\textbf{AI binary classifier.} 
Following the idea of adversarial generative models, we trained an AI classifier to check whether the synthetic tumors could confuse deep learning algorithms. A ResNet-50~\cite{he2016identity} was trained to classify extracted 2D tumors with annotation masks.
As shown in \figureautorefname~\ref{fig:roc_human_study}, the trained ResNet-50 showed a similar classification ability to the senior radiologist: it could classify all sizes of tumors better than small tumors. This is evidence that our synthetic tumor algorithm works better on small tumors, making it a promising method to use to enhance the prevalence of small tumors in pancreatic CT datasets.

\smallskip\noindent\textbf{Results on the \texttt{JHH} dataset.} We evaluated the synthetic tumors using an in-house dataset. Following the state-of-the-art work on pancreatic tumor segmentation~\cite{xia2022felix}, we trained a single-phase segmentation neural network with the nnU-Net framework~\cite{isensee2021nnu}. This network training strategy was applied to all the following experiments. 
\tableautorefname~\ref{tab1e-FROC} shows that models trained with synthetic tumors for Cyst and PDAC could achieve a similar tumor detection performance. For PDAC, AI trained on synthetic tumors outperformed the model trained by real data significantly. Actually, models trained with synthetic data could achieve a lower FP/subject at 0.4 and still perform a reasonable sensitivity (around 95\%). 
To test whether our data could help boost the performance of small tumor detection, we stratified the detection results based on the radius of the tumors, as shown in \figureautorefname~\ref{fig:detection}. Tumors with a radius $<20mm$ are considered small tumors. Because in the training stage, we synthesized more small tumors than their actual prevalence, model trained with synthetic data has a better ability to detect small tumors. However, with extremely small ones (radius $<10mm$), further research is still needed. But it will be a challenging task since a small pancreatic tumor looks might exactly like its neighbor and could be hard to detect with only one phase of CT.

\smallskip\noindent\textbf{Results on the two public datasets.} 
On these two public datasets, we showed that a hybrid data training strategy (to train the model with both real data and synthetic data) could provide a much stronger performance. All cases in these two public datasets are diseased, but for \texttt{MSD}, the subtype of tumor is not specified. We adopted our tumor generation method directly and ignored the fact that our synthetic tumor could overlap with the real tumor. \tableautorefname~\ref{tab1e-hybrid} shows that segmentation tasks could benefit from involving synthetic data in the training loop, as the Dice scores both improved from training with real data only.

\smallskip\noindent\textbf{Domain adaptation from public to in-house datasets.} 
Finally, we tested models trained with public datasets on our in-house \texttt{JHH} dataset. \tableautorefname~\ref{tab1e-ablation} shows that with synthetic data, models could perform better on an unforeseen dataset. However, we need to notice that the model generation ability is still limited by the similarities between the source dataset and the target dataset. For example, \texttt{PCL} has no PDAC, so even if we augmented the dataset with synthetic PDAC, its ability to detect PDAC on \texttt{JHH} was still below expectation.

\section{Discussion \& Conclusion}
\label{sec:discussion_conclusion}

In this paper, we presented a method to generate pancreatic tumors with label-free tumor synthesis, and these synthetic tumors could help boost the performance of detection and localization of pancreatic cancer. Compared with synthetic liver tumors, synthesizing pancreatic tumors is a more difficult task: the size of the organ is small, and the contrast of the tumor with its surrounding tissues is subtle. And thus, the generation of synthetic tumors could help the training of deep neural networks more, since the networks could be trained with a larger variety of data.
We provided evidence that training with synthetic data and hybrid training could boost the segmentation performance of pancreatic tumors. We still observed the weakness of all current methods: the detection rate is still low for small pancreatic tumors.
This version of the synthetic method depends on some key parameters, for example, differences in intensities and size. With related clinical knowledge, we could generate these kinds of parameters for other kinds of tumors in different organs. On the other hand, it would be better if such parameters could also be learned in a zero-shot way.
Future potential research for synthetic pancreatic tumors could include (I) synthesizing Pancreatic neuroendocrine tumors (pNET), since pNET is a more difficult tumor to identify and diagnose and (II) automatic parameter learning with a generative adversarial network or diffusion model. 

In conclusion, synthetic tumors provide a promising direction to generate large-scale, label-free synthetic data and annotations for training AI in organs, such as the pancreas, that lack detailed, per-voxel tumor annotations. The use of synthetic tumors is attractive as it requires no manual annotation and enables us to specify the location, size, shape, intensity, and texture of the tumor to be synthesized. Moreover, synthetic tumors can also be used as supplementary data when training AI with real tumors. Specifically, a variety of small tumors automatically generated by our method has demonstrated the potential to improve the early detection of pancreatic cancer.

\medskip\noindent\textbf{Acknowledgements.} This work was supported by the Lustgarten Foundation for Pancreatic Cancer Research and the Patrick J. McGovern Foundation Award. We appreciate the effort of the nnU-Net Team~\cite{isensee2021nnu} to provide open-source code for the community. We thank Qixin Hu, Yuxiang Lai, and Wenxuan Li for their constructive suggestions at several stages of the project.

%
%
%
\newpage
\bibliographystyle{splncs04}
\bibliography{references,zzhou}

\end{document}